# A Traffic Incident Management Framework for Vehicular Ad Hoc Networks

Rezvi Shahariar [1] and Chris Phillips [2]

[1] Institute of Information Technology, University of Dhaka, Dhaka, Bangladesh
[2] School of Electronic Engineering and Computer Science, Queen Mary, University of London, England

## Abstract

*Vehicular Ad Hoc Networks (VANETs) support the information dissemination among vehicles, Roadside Units (RSUs), and a Trust Authority (TA). A trust model evaluates an entity or data or both to determine truthfulness. A security model confirms authentication, integrity, availability, non repudiation issues. With these aspects in mind, many models have been proposed in literature. Furthermore, many information dissemination approaches are proposed. However, the lack of a model that can manage traffic incidents completely inspires this work. This paper details how and when a message needs to be generated and relayed so that the incidents can be reported and managed in a timely manner. This paper addresses this challenge by providing a traffic incident management model to manage several traffic incidents efficiently. Additionally, we simulate this model using the VEINS simulator with vehicles, RSUs, and a TA. From the experiments, we measure the average number of transmissions required for reporting a single traffic incident while varying the vehicle density and relaying considerations. We consider two types of relaying. In one series of experiments, messages from regular vehicles and RSUs are relayed up to four hops. In another series of experiments, messages from the regular vehicles and RSUs are relayed until their generation time reaches sixty seconds. Additionally, messages from the official vehicles are relayed when they approach an incident or when the incident is cleared. Results from the simulations show that more vehicles are informed with four-hop relaying than sixty-second relaying in both cases.*

## 1. Introduction

A Vehicular Ad Hoc Network (VANET) consists of vehicles, Roadside Units (RSUs), and a Trust Authority (TA) which allows emergency and traffic announcements [1]. There may be more than one TA which provides registration and deregistration of drivers or RSUs. A typical arrangement of a VANET is illustrated in Figure 1.It shows a direct connection between RSU-RSU and RSU-TA. An RSU announces messages within its coverage area. Thus, many RSUs are needed to serve users on roads. Vehicles run on roads and RSUs are placed alongside the roads. Many types of vehicles run on roads for instance: regular cars, taxis, buses, trucks, ambulances, police vehicles, and so on. They obtain traffic updates from RSUs which they also relay to neighbouring vehicles. These communications continue until issues are resolved or meet the conditions for stopping announcements. RSUs themselves also send messages to instruct nearby RSUs to announce the same information which helps vehicles in nearby regions to choose alternate routes to reach their destination safely. RSUs also communicate with official vehicles (police, ambulance, fire service vehicles)when approaching an event or resolving an incident. This sort of communication is more trustworthy as official vehicles are considered reliable.

Trustworthy announcements are required to achieve driving comfort as suggested by [1]. In this paper,the effect of the false announcements of an event is examined which results in drivers needing additional travel time to reach their destination. A VANET is used in other applications as well, for example, collecting tolls, finding nearby services (petrol pump/restaurant), and so on [2].

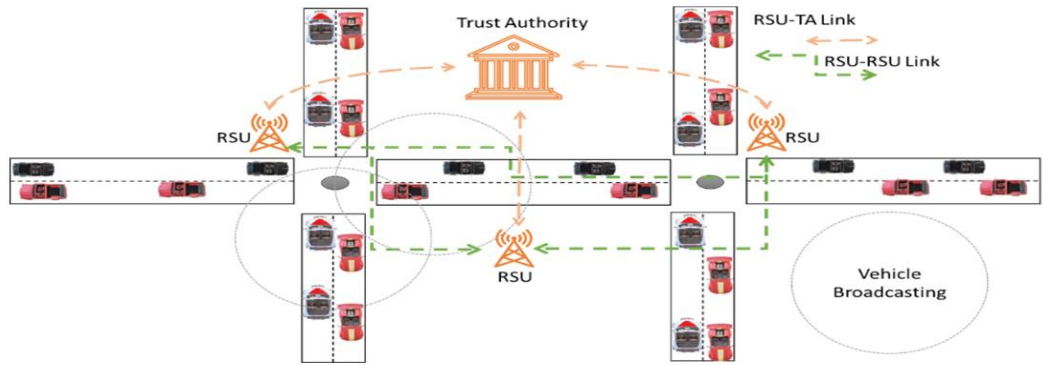

Figure 1. A Typical VANET Example

In vehicular communication, a vehicle announces an event whenever it observes it first. Sometimes, this announcement is generated from multiple vehicles. Based on the announcement, each vehicle takes a decision to avoid the traffic incident. After an announcement in the network, the authority should send different traffic updates to users. For example, whenever they start handling the event, or when the event is sorted. When an event is resolved, users need to be communicated in a timely manner. For example, when a road is cleared, an RSU needs to spread this information to nearby RSUs so that vehicles from nearby regions may reuse the road. Otherwise, vehicles from neighbouring regions would avoid the road. Thus, event handling means announcing messages including when to avoid and when to use a road. Besides, some roads are congested whilst others are free to use. This should be communicated dynamically so that road users in different regions travel safely and face minimal congestion.

There are some management activities to consider when an incident arises. For instance, when an accident occurs, upon receiving announcements, both vehicles and RSUs relay the message until specific conditions are met. Besides, the authority instructs an RSU to inform other RSUs to announce the same information across a wider area. Also, whenever the incident is resolved, then this needs to be announced so vehicles can safely use the road again. To the best of our knowledge, there is no single existing trust/security/information dissemination model which provides this type of communication or message forwarding sequence for resolving traffic incidents in VANETs. Hence, this work proposes several sequences of messages for different traffic incidents. These messages are circulated around the event zone where an incident took place. The sequences of messages are different for different traffic scenarios. For this, we have considered some well-known traffic events, for instance, accidents, traffic jams, congestion, obstacles, stranded vehicles, service discovery, floods/debris on roads, and traffic element problems. We realise these are not a complete list of incidents that may happen in a VANET but represent initial examples. Later, one can develop further sequences of messages to solve other identifiable traffic incidents.

In the literature, many models exist that consider various traffic incidents but lack the message-forwarding sequences for specific traffic scenarios. It is necessary to develop a model which delineates a procedure for incident announcement and resolution. This also includes the relaying of the same event multiple times by different entities, as needed. This helps to free up roads near the incident so that official vehicles can reach the event to sort it. Also, this model can announce when a road is available to use again. We believe this can be used as a communication model for VANETs where a trust model evaluates the event announcement, a security mechanism enforces the security, and it defines different control messages and their sequence of announcements. These announcements are necessary to inform vehicles about the current situation. Vehicles can follow the instructions from the RSUs to avoid issues. Relaying and controlled re announcements are necessary with this model to reach incoming vehicles as the network is dynamic. Official vehicles maybe near or parked at a specific location so that they can be requested to investigate incidents. If they are far away from a specific incident, they can be reached using the interconnected RSU network. As they are official, their participation is required to verify an incident. While they are resolving an incident, they may need to communicate with a RSU, for example, sending an "avoid road" message. Clearly, this improves traffic incident management. The proposed communication framework delineates the overall process of traffic incident management. This define show and when a specific message needs to be announced for resolving a certain traffic incident. In this work, the following contributions are made:

- First, it proposes an incident management model to be used with a trust or security scheme or with a traffic condition dissemination system considering different road traffic scenarios. In this model, we have considered some well-known traffic events, for instance, announcements and management of accidents, traffic jams, congestion, obstacles, stranded vehicles, service discovery (nearby parking, Wi-Fi, Patrol pump), floods/debris on road, and traffic element problems. We develop different sequences of messages for different incidents. This helps users experience better traffic comfort on roads. For example, when resolving an accident, we use an accident announcement, and official vehicle messages to alert incidents and state when they are resolved. This series of messages instructs users to take appropriate actions, for instance, when to use or when not to use a specific section of a road. Thus, better road safety and traffic comfort is achieved using this model.

- Second, we implement these scenarios in VEINS. We measure the average communication overhead for accident announcements considering different relaying conditions. For example, we try to reach vehicles at four hops distance in one set of experiments, and in another, we relay the same event with the message freshness set to 60seconds. The average communication overhead for each scenario is compared considering 4-hop versus the60-second relaying alternative. Results suggest that more vehicles can be informed with the 4-hop relaying approach than with the 60-second relaying.

Section 1encompassesan introduction to VANETs and states why we need a traffic incident management model. Section 2reviews existing models which consider the dissemination of messages or managing trust or handling security issues in VANETs. In Section 3, we explain the detailed process for managing different traffic incidents in VANETs along with sequence diagrams showing the order of message generation. Section 4then provides implementation details and includes a relaying-based performance comparison of the communication overhead based on the proposed traffic management process. Finally, Section 5concludes this work by highlighting what we have achieved with this framework so far. This also includes some suggestions for future work relating to this topic.

## 2. REVIEW OF EXISTING TRUST, SECURITY, AND TRAFFIC INFORMATION DISSEMINATION MODEL

As this model is proposed for traffic information dissemination, it can be used with a trust/security model or as an information dissemination model for resolving traffic incidents. Here we highlight some well-known trust, security, and information dissemination models based on different technologies, for example, blockchain, artificial intelligence, machine learning, and fuzzy logic-based communication. To the best of our knowledge, no single model exists which delineates the message forwarding sequences similar to our proposed model. Some existing trust models evaluate the trust of vehicles, others evaluate only data from vehicles. Also, some models evaluate both the trust of vehicles and the data from vehicles. Security models address some known security aspects which are authentication, privacy, integrity, and non repudiation. Furthermore, some traffic update models present work to disseminate traffic information to users. We highlight some well-known trust/security/information dissemination-based models to review the extent of their work and their limitations.

In [3], the authors provide a trust-oriented message evaluation system incorporating a game-based reward or punishment scheme for encouraging trustworthy messages from vehicles. It uses blockchain to maintain a consistent trust score in VANETs. The analysis model confirms vehicle participation and the thwarting ability of false information. Nonetheless, when a traffic incident occurs, it does not provide message-forwarding sequences to manage different traffic incidents. In [4], the researchers propose a direct and indirect trust-based model to prevent false information sharing among the vehicles. Direct trust is calculated using a Bayesian classifier and indirect trust is computed from the active detection-based method. A blockchain stores the tamper-proof indirect trust data. Nevertheless, this scheme lacks a traffic management module in terms of defined message-forwarding sequences for resolving different incidents. Reference [5] proposes a trust model considering the transmission path of messages as an important metric for trust evaluation besides direct and indirect trust. This uses the Dempster-Shafer Theorem (DST) to fuse these metrics. Also, path-backtracking mechanisms, based on a message transmission path, detect malicious behaviour. This approach introduces a new metric but does not provide a traffic management model for VANETs. In [6], a direct and indirect trust-based evaluation model is presented which uses different machine learning models (for example, multilayer perceptron, support vector machine, random forest, decision tree and KNN) to identify suspicious behaviour by analysing historical trust interactions among vehicles. Even so, this does not provide any message forwarding sequence to manage the traffic incident completely.

[7] proposes a trust-cascading event dissemination model which embeds entity trust in the data trust computation. Here, entity trust is used as an important weight to find the data trust. This model considers various stages, for example, registration of vehicles and RSUs, trust certificate queries, feedback reporting, trust revising, and vehicle revocation. Nonetheless, there is no traffic incident model which delineate show an event can be sorted. In [8], the authors propose a trust rank algorithm which employs both local and global trust computation of vehicles. Local trust is computed using Bayesian Inference. After that, some seed vehicles are selected using local trust and social factors. Next, global trust is computed using the reputed seed vehicles and the local trust structure (trust link graph). Even so, it does not provide a traffic incident management model for resolving road traffic situations. Reference [9] proposes a three-factor (expectation, risk, and confidence value) based data trust model which relates trust with these factors. The model selects the beacon message with high expectation, low risk, and high confidence. Nevertheless, this model does not suggest an approach to manage incidents by including message forwarding sequences from RSUs, authoritative, and regular vehicles.

In [10], an adaptive trust model incorporating a reinforcement learning-based framework is presented to capture the behaviour of neighbouring vehicles. This model achieves trustworthy communication by considering trust and link-lifetime to select a good neighbour from the neighbourhood. Nonetheless, the total process depicting how to recover from an incident is not included in the model. In [11], to address conflicting reports about a road incident, a risk-based trust model is given. This model comes with a decision-making process by estimating the risk of alternative actions and then selecting the report with the lowest risk. Their analysis shows that the risk-based approach outperforms the trust-based method. Even so, this does not come with message forwarding sequences considering different entities within a VANET to resolve traffic incidents. [12] presents a multidimensional trust model using direct and indirect trust but it does not include any traffic incident management module to address specific incidents. In [13], a trust model is presented which prevents the selection of a malicious vehicle as the cluster head. A trust value is associated with a vehicle's available resources to be used as the cluster head and the proxy cluster head. Likewise, it does not present an incident management module to handle road issues.

In [14],local trust and global trust are computed using Bayes Trust and vehicle rank algorithms. When enough local trust information exists, this model can identify malicious and benevolent vehicles. Nonetheless, this trust model does not include a process to recover from a traffic incident by sending messages from different entities in a specified manner. Reference [15] uses cognition to learn from reports of the surroundings and prepare a context around an event. Additionally, it includes a malicious vehicle detection and isolation mechanism. However complete recovery from an incident is not given in this model. [16] also presents a context-aware based trust model which evaluates the trust of events to limit the influence of false information in the driving decision-making process. In this model, trust evaluation depends on the availability of information, current strategy, and the decision is not affected by conflicting reports and entity trust. A reinforcement learning model is also included to adjust the evaluation strategy for different traffic scenarios. Nevertheless, a recovering process from the incident illustrating the message-forwarding sequences is absent. In [17], a trust model is presented which separates malicious vehicles using a tier-based message dissemination approach. Each vehicle verifies the authenticity of the event to issue a trust value to the source of the event. A recovery strategy detailing the different messages generated from various entities in the network is absent.

[18] uses DST to estimate uncertainty in VANETs using both direct and indirect trust scores. While computing the final trust score, this model uses penalty, forgetting, reward, and uncertainty-based factors. Nonetheless, it does not present a traffic incident management model to resolve the incident. This type of model is required as drivers need to decide their next movement along the road. Thus, without road clearance assurance, drivers may drive through the region where an incident has occurred. In [19], the trust model considers deploying fog nodes for event detection, cluster head selection, and misbehaviour detection. This also presents a task-based trust model to evaluate the trust of a vehicle based on the type of tasks it requests. Nevertheless, it does not provide a sequence of messages to follow to recover from different traffic incidents. [20] proposes active detection and blockchain-based trust model. Active detection is used to identify dishonest vehicles and blockchain is used to maintain the consistency and tamper-proofness of the trust data. Even so, a complete sequence of messages to resolve specific traffic incidents on the road is not given. In [21], a multilevel trust model is proposed to manage traffic effectively. This model defines trust by detection of trust, reference trust, transmission trust, and stability trust. Once again, it does not define the sequence of messages communication would achieve to manage traffic incidents effectively.

In [22], a sender-side evaluation-based trust model is presented to achieve communication efficiency in terms of low communication overhead and delay. This is based on deploying a Tamper-Proof Device (TPD) in every vehicle to manage the trust of the sender vehicle securely. A sender's TPD allocates trust based on reward and punishment. Punishment and reward are determined using accuracy, responsiveness, and travel distance from the incident location. An RSU also rewards or punishes a driver based on win or lose disputes using a fixed RSU reward and punishment strategy. In [23], an amendment of the reward or punishment strategy from the model in [22] is presented using fuzzy logic. Fuzzy logic considers driver history, feedback provider support to sender/reporter drivers, and incident severity to assign an appropriate level of reward or punishment to the concerned driver. Even so, both works lack the presentation of message-forwarding sequences to manage different kinds of incidents in VANETs.

In [24], the Identity-based Batch Verification (IBV) model proposes to achieve security and communication efficiency for VANETs. Considering the spreading of false information and privacy issues, this model achieves anonymous authentication, integrity, privacy/confidentiality, and traceability. As with the trust models stated earlier, this model does not illustrate a complete sequence of messages to recover from a traffic incident.[25] highlights the security requirements for VANETs which are source authentication of messages, integrity of messages, non-repudiation, maintaining timing requirements, access control, privacy of messages, anonymous messaging and traceability of messages when required, denial of service which leads to reduced quality of service and so on. Nonetheless, there is no detailed process to manage an incident from its happening to clearing the road.

In [26], a PASTA threat model is developed to thwart adversarial AI threats. In this work, some known AI attacks, vulnerabilities, and mitigation policies are stated. Additionally, risk and the impact of security attacks are discussed, and hints are given for future development. Still, it lacks complete management of security incidents in VANETs. In [27], a CPPA signature scheme based on Elliptical Curve Cryptography (ECC) is used to provide ultra-low transmission delays and to update the secure secret key in the TPD of an unrevoked vehicle using a pseudo-random function and Shamir's secret key algorithm. Even so, it does not show a complete sequence of messages to recover from a traffic incident.

In [28], a multi-agent based safety information propagation approach is presented using the distance of a vehicle in the exterior region. This model selects the vehicle to reduce the transmission delay. This model considers accidents, extreme fog, bad roads, heavy rain, extreme congestion, and medical/toll/fire/map assistance as traffic incidents. Nonetheless, it lacks a complete sequence of messages to manage traffic incidents. Furthermore, [29] proposes a clustering approach to achieve data dissemination in VANETs. The authors find the network is more stable with this approach by reaching all vehicles in the simulation. Nevertheless, a complete sequence of messages is missing to guide the process of resolving incidents in VANETs. In [30], a policy-based framework is designed to support data dissemination in VANETs. RSUs manage access policy, combine multiple policies, and also provide policy conflict resolution to handle policy conflicts between an RSU and a vehicle. Nevertheless, it lacks a traffic incident management framework for VANETs to resolve specific incidents.

## 3. A TRAFFIC INCIDENT MANAGEMENT FRAMEWORK

This section defines the sequences of messages for selected sets of traffic events and delineates the process of traffic management to resolve road issues. We have considered accidents, traffic jams/congestions, obstacles, obstacle clearances, diversions, debris on the road, service discovery, road defects, and traffic element malfunctions. Next, each incident is illustrated, and a sequence diagram is provided to show the message generation sequence in the network.

## 3.1. Announcement of an Accident

Imagine an accident happens on a road called "X," and a vehicle notices it whilst traveling on that same road. The vehicle sends an accident notification to other vehicles and RSUs. Upon receiving this message from intermediate vehicles via relaying, an RSU then retransmits the same notification within the VANET. The RSU regularly broadcasts an "avoid road" message due to the unavailability of the road "X." It also coordinates with nearby RSUs to spread the "accident on road X" message, helping prevent potential traffic congestion near the incident. Vehicles receiving this alert from any source will avoid road "X." At the same time if a police car receives the message, it will repeatedly broadcast "attending road X" and "clear road X " notifications to quickly reach the accident site. If an RSU receives a message from a police car, it will announce a "restricted movement on road X" notification. Whenever an RSU gets such a message from a police car, it periodically broadcasts an alert stating, "Only police, ambulance, or fire service vehicles are allowed on road X with the highest priority, and all other vehicles must take an alternate route." This raises awareness among the nearby drivers, prompting regular vehicles to avoid road X, allowing official vehicles to reach the incident site more quickly. As a result, the situation can be resolved faster. Once the incident is cleared, the police vehicle sends a "sorted road X " message. Finally, the RSU broadcasts this "cleared road X" status message multiple times at regular intervals, allowing vehicles to resume using road X.

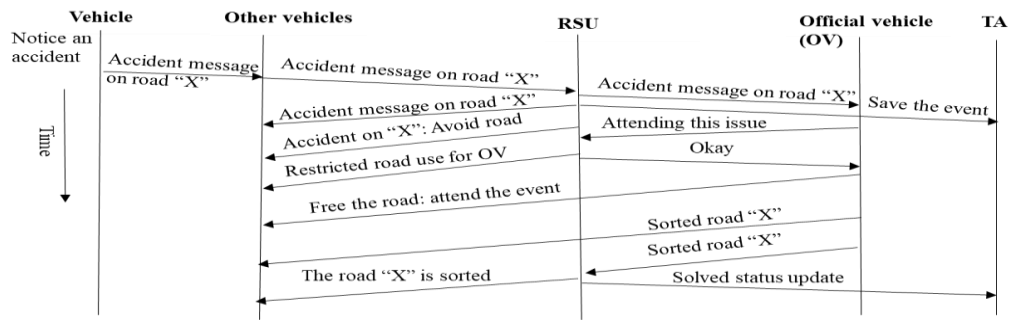

Figure 2. Sequence Diagram for Announcing an Accident Event

The RSU also forwards the "cleared road X" message to nearby RSUs, which then broadcast the message multiple times. This is how the proposed incident management framework addresses an accident event. "Attending road X" indicates that an official vehicle is trying to investigate the event, while "cleared road X" signals that the incident on road X has been cleared. The name of each message reflects the nature of the event for its recipients. This applies to all network events considered with this framework. Figure 2 shows a sequence diagram for broadcasting an accident event. This diagram states what are the different messages we require for accident event handling, and in which order they need to be shared among various entities. Figure 3 demonstrates the entire process of handling an accident event, while Figure 4 outlines the steps involved in resolving the accident.

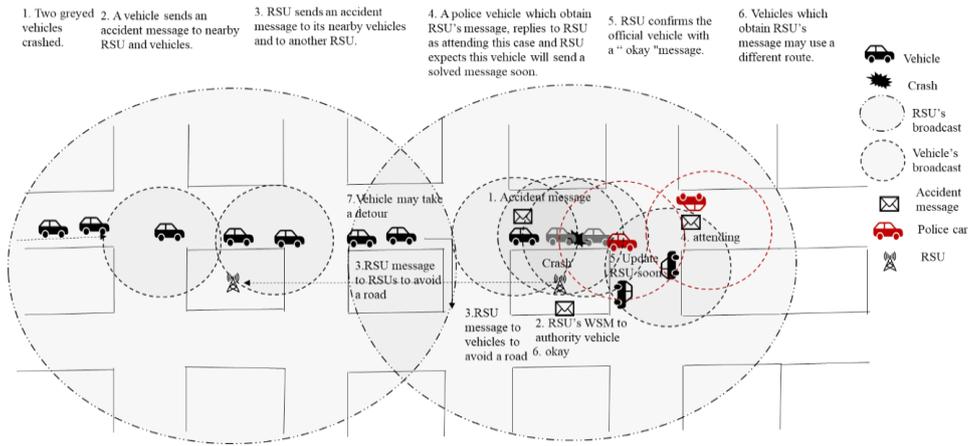

Figure 3. Announcement of an Accident Message

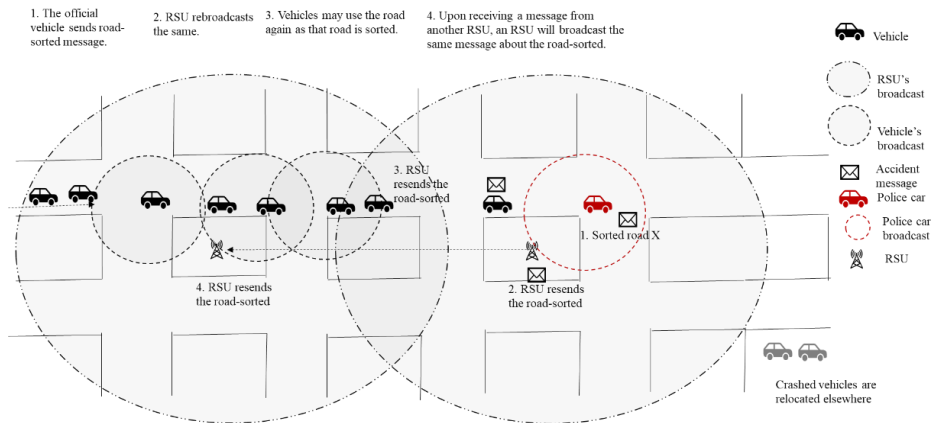

Figure 4. Scenario of a Police Car Managing an Accident Incident

An official vehicle may receive an accident message from a vehicle before an RSU. In this case, the official vehicle initiates the recovery process. The official vehicle that receives the message first, retransmits it. If a nearby RSU receives it, it periodically broadcasts the accident message within its range to alert other vehicles. Once the incident is resolved, the police car sends a "cleared road X" message, which may reach an RSU via intermediate vehicles. This RSU then informs other nearby RSUs, and they also periodically broadcast the traffic update. This is illustrated in Figures 5, 6, and 7.

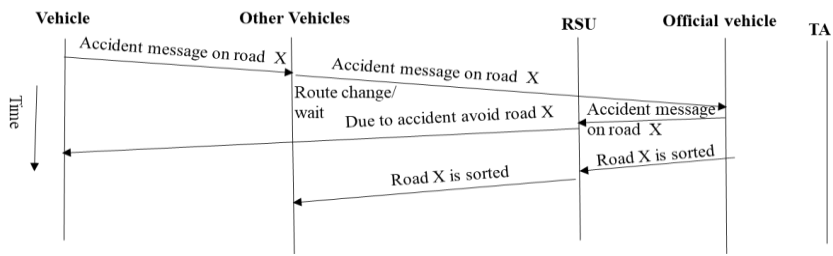

Figure 5. Sequence Diagram for Handling an Accident Message by a Police Car

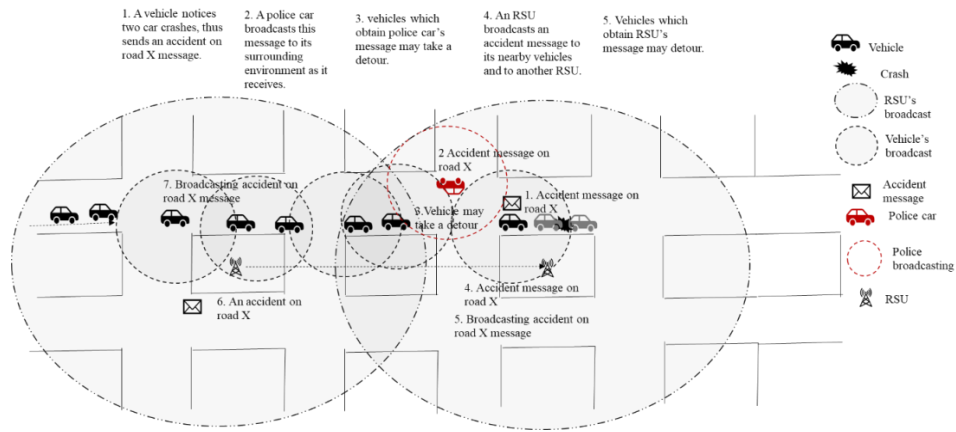

Figure 6. Scenario of a Police Car Receiving an Accident Message Before an RSU

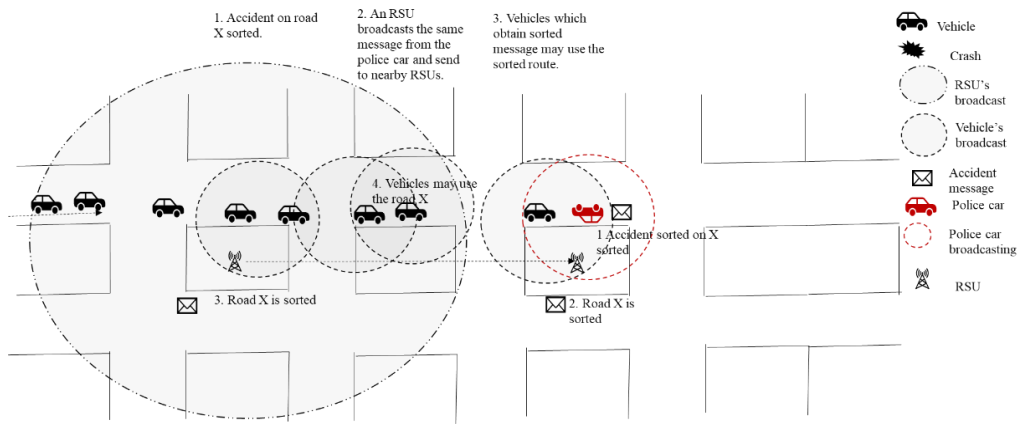

Figure 7.Scenario of a Police Car Resolving an Accident Event

## 3.2. Announcement of a Traffic Jam and Congestion

A vehicle broadcasts a "traffic jam on road X" message when its speed drops below 0.1 m/s and it has been stationary for over 30 seconds while noticing other vehicles queued ahead. This message is relayed by intermediate vehicles and may eventually reach an RSU. The RSU periodically retransmits the "traffic jam on road X" message to vehicles within its range until it receives a traffic-clear "cleared road X" message. The RSU also sends the "traffic jam on road X" message to nearby RSUs for further distribution across the VANET. This helps prevent more severe traffic jams by notifying drivers before the situation worsens. Vehicles receiving the message may either detour or remain in the queue. Later, when a vehicle detects that the road is clear, it broadcasts a "road clear" message. An RSU then rebroadcasts the "cleared road" message multiple times. The RSU that initially sent the traffic jam message now announces the road is clear and forwards this update to nearby RSUs. Vehicles receiving the traffic update can then use the road "X" again. Additionally, a different congestion message is generated when a vehicle's speed is between 1-13 m/s for 60 to 90 seconds, and this message is processed similarly to the traffic jam message. Figures 8 and 9 illustrate the message sequence diagrams for traffic jam and congestion, while Figures 10 and 11 show the corresponding scenario diagrams for broadcasting and resolving these incidents.

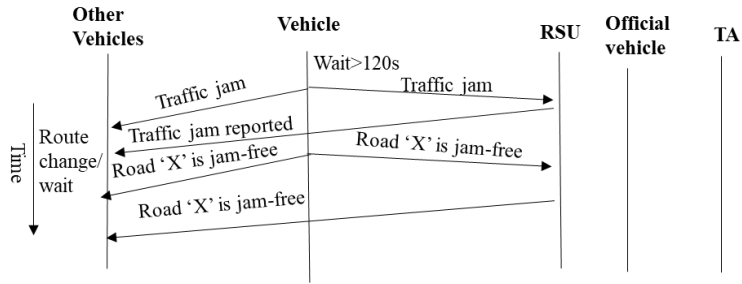

Figure 8. Sequence Diagram for Announcing a Traffic Jam

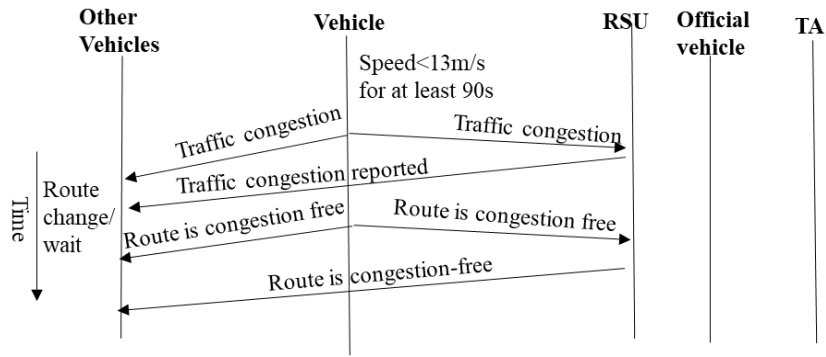

Figure 9. Sequence Diagram for announcing a Traffic Congestion

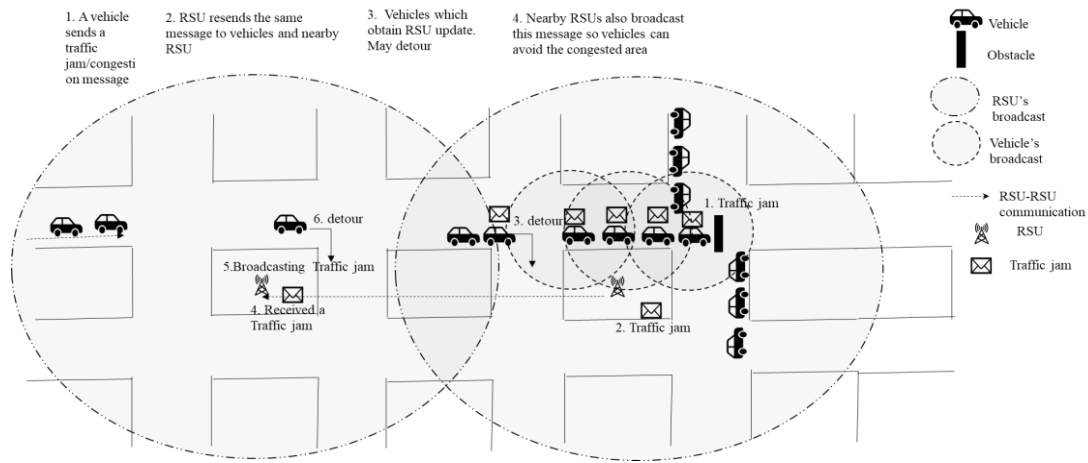

Figure 10. Scenario for Announcing a Traffic Jam/Congestion

### 3.3. Announcement of Obstacles and their Clearance

A vehicle detects an obstacle on road X and sends a message stating, "Obstacle on road X." This message is passed along to an RSU via intermediate vehicles. The RSU then retransmits the message to other vehicles within its coverage area and also notifies nearby RSUs. By this time, an official vehicle may receive the message and respond to the RSU with an "attending road X" update. Other RSUs continue to broadcast the traffic alert within their range. Vehicles that receive this update can choose to take an alternate route if they prefer. Later, once the police car arrives and removes the obstruction, it sends a message to the RSU saying, "No obstacles on road X." The RSU then relays the message to surrounding RSUs. These RSUs broadcast the update

28

that the obstacle on road X has been cleared, allowing nearby vehicles to safely use the road again. Figure 12 illustrates the sequence diagram for this scenario, while Figures 13 and 14 show the corresponding visual depiction of the situation.

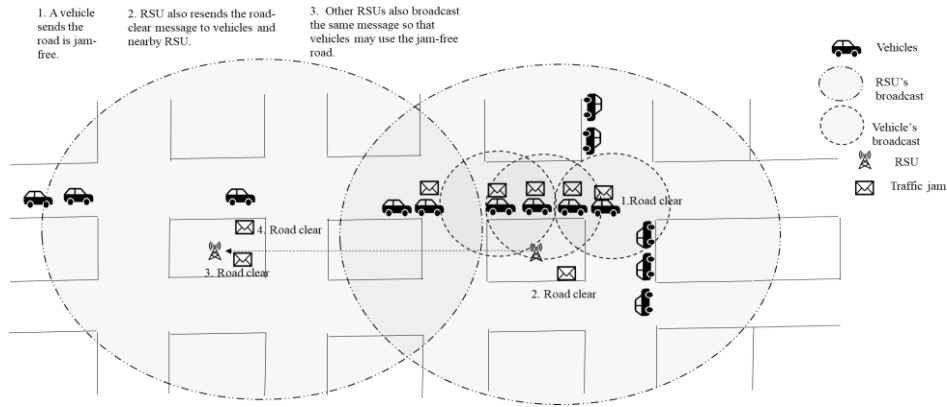

Figure 11. Scenario of Reporting a Traffic Jam/Congestion Clear Message

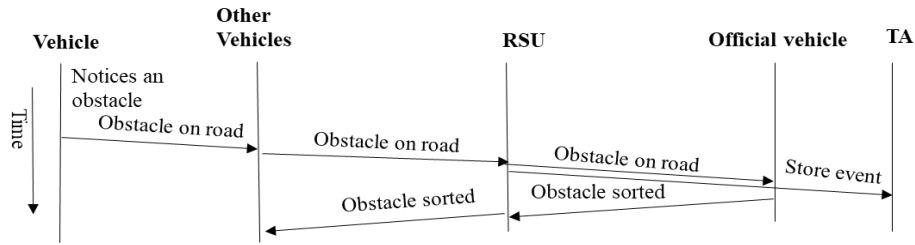

Figure12. Sequence Diagram forAnnouncing an Obstacle on Road Message

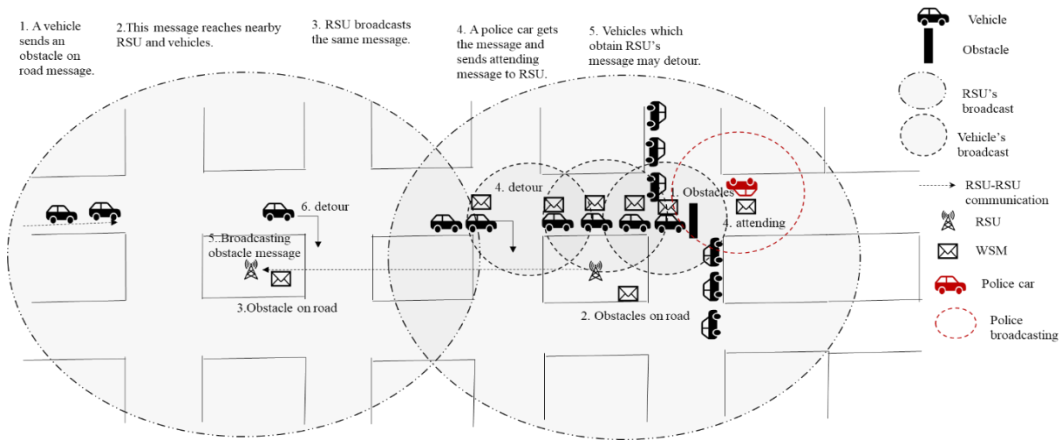

Figure 13. Scenario for Reporting an Obstacle on Road Message

### 3.4. Announcement of a Diversion

An official vehicle sends a "diversion on road Y" message when the authorities wish to prevent vehicles from using a particular road. Upon receiving this message, an RSU notifies nearby RSUs. These RSUs periodically transmit the same message to vehicles within their coverage area. Vehicles that receive this update can choose an alternate route to their destination without

29

International Journal of Wireless & Mobile Networks (IJWMN), Vol.17, No. 2, April 2025

any issues. Figure 15 illustrates the sequence diagram for the diversion scenario, while Figure 16 shows the process of announcing the road diversion in the VANET.

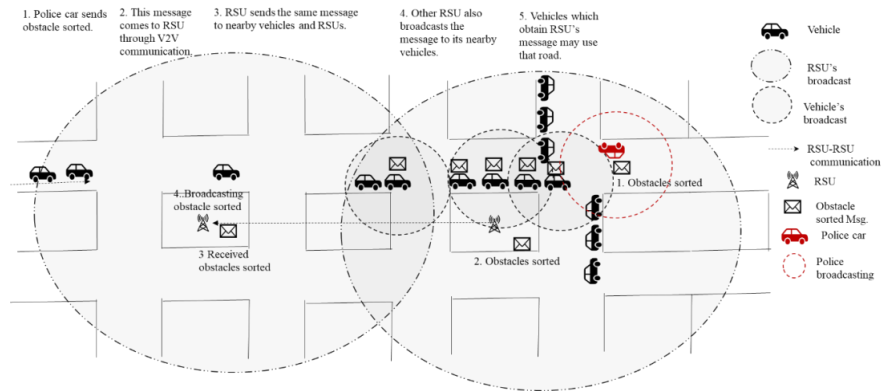

Figure 14. Scenario for Reporting an Obstacle on Road Clear Message

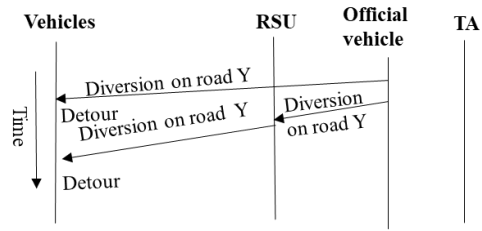

Figure 15. Sequence Diagram for Announcing a Diversion on Road Y

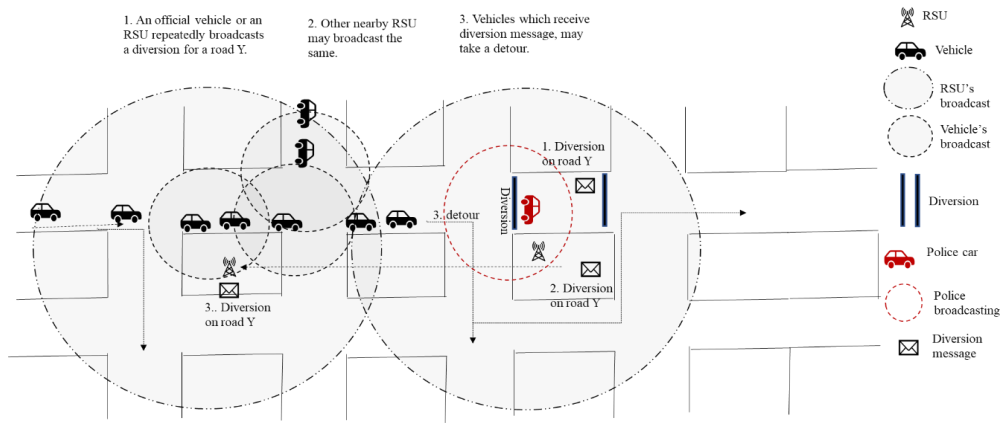

Figure 16. Process of Reporting a Diversion Message on Road Y

### 3.5. Announcement of a Stranded Vehicle

When a vehicle detects a stranded vehicle on road X, it broadcasts a message saying "stranded vehicle on road X" to other vehicles and the RSU. The RSU then forwards this message to the police car and other nearby RSUs. Upon receiving the message, the police car responds, indicating that it is attending to the situation. Meanwhile, the RSU continues to rebroadcast the message. Later, the RSU receives a "cleared road X" message from the police car, which had previously addressed the issue. The RSU then retransmits this "cleared road" message to the vehicles within its range and notifies nearby RSUs. Other nearby RSUs also broadcast the "cleared road" message to the vehicles in their transmission areas. Figure 17 illustrates the sequence diagram for a stranded vehicle message on the road. Figures 18 and 19 depict the process of broadcasting the stranded vehicle message within the VANET.

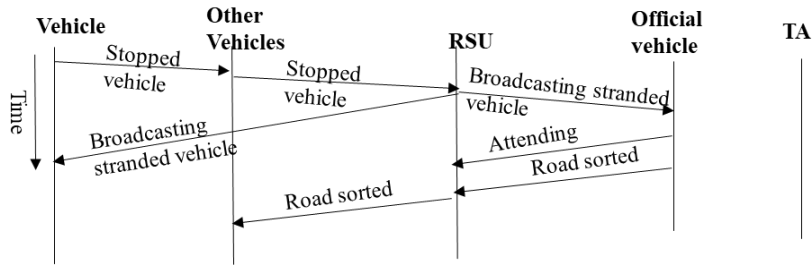

Figure 17. Sequence Diagram of Broadcasting a Stranded Vehicle Message on a Road

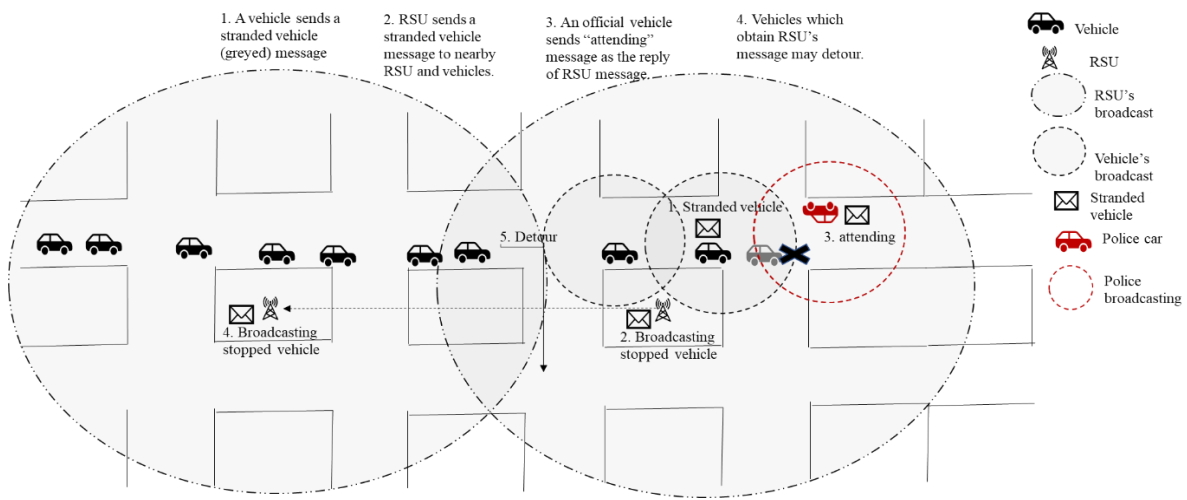

Figure 18. Process of Broadcasting a Stranded Vehicle Message on Road X

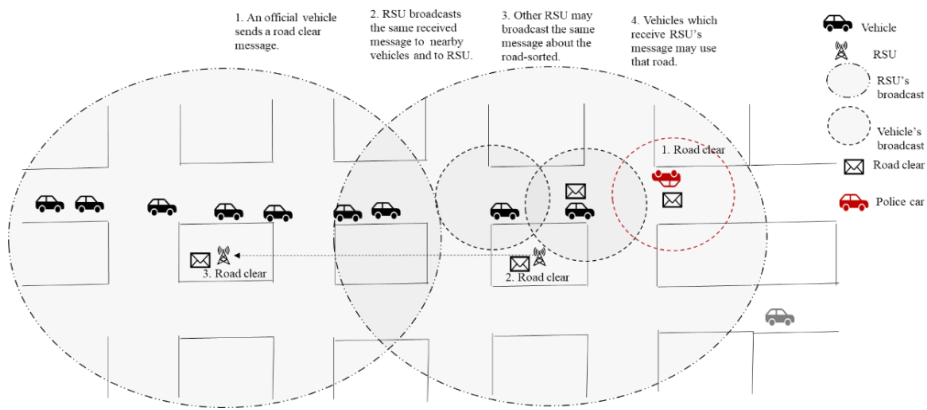

Figure 19. Process of Announcing Stranded Vehicle Clear Message on Road X

### 3.6. Debris on Road

When a vehicle detects debris on road X, it broadcasts a message about the obstruction. The RSU forwards this information directly to the Traffic Authority (TA), which then coordinates with the

31

RTA to send official personnel to address the issue. Once the debris is cleared, the TA sends a message to the RSU stating "resolved: debris on road X." The RSU periodically retransmits this message within its range. As a result, all vehicles within the coverage area are notified and can safely use the road again. Figure 20 illustrates the sequence diagram for debris on the road. Figure 21 depicts the process of broadcasting the debris message on the road.

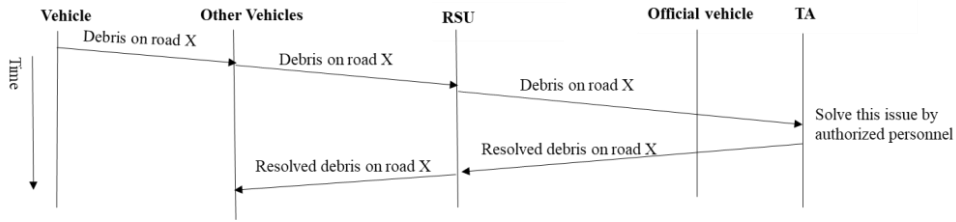

Figure 20. Sequence Diagram for Reporting a Debris on Road Message

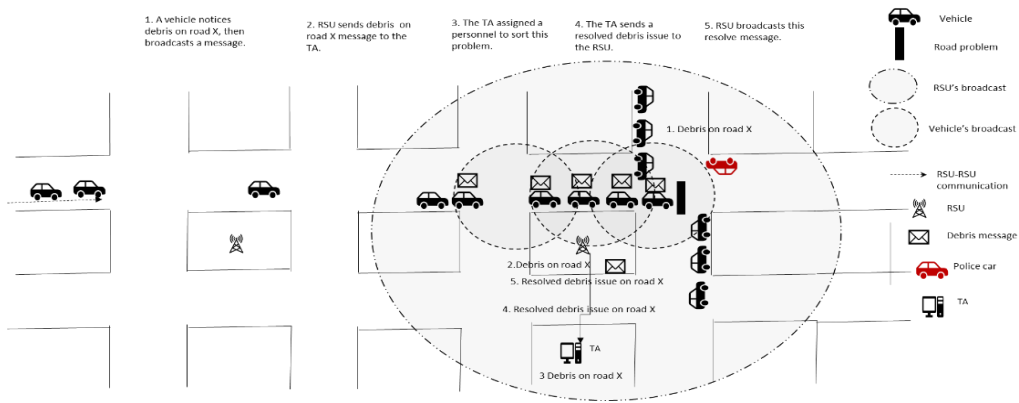

Figure 21. Process of Resolving Debris on a Road

## 3.7. Service Discovery from a Regular Vehicle

Imagine a vehicle searching for the location of a nearby petrol pump and requesting assistance from an RSU by sending the message, "Do you know a nearby petrol pump?" This message passes through intermediate vehicles and reaches a nearby RSU. In response, the RSU sends back the information, "The location of a nearby petrol pump is on road X." The vehicle then uses this information. Figure 22 illustrates the sequence diagram for a general service lookup in a VANET, while Figure 23 shows the sequence diagram for locating a petrol pump in a VANET. Figure 24 depicts the process of searching for a petrol pump within the VANET. Using this process, a vehicle can look up any registered service in the VANET, for example, parking area, restaurant, Internet facility, and so on.

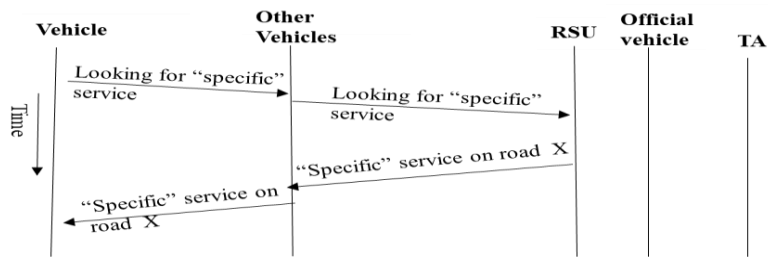

Figure 22. Sequence Diagram to Discover a Service

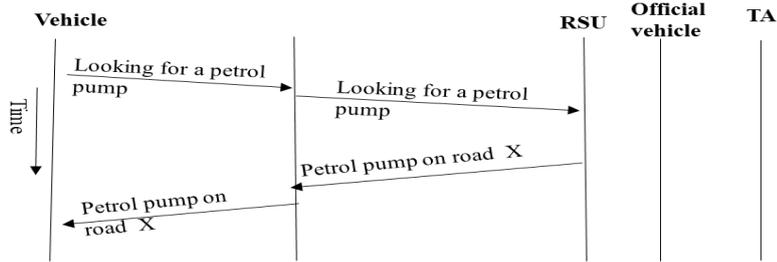

Figure 23. Sequence Diagram to Discover a Petrol Pump

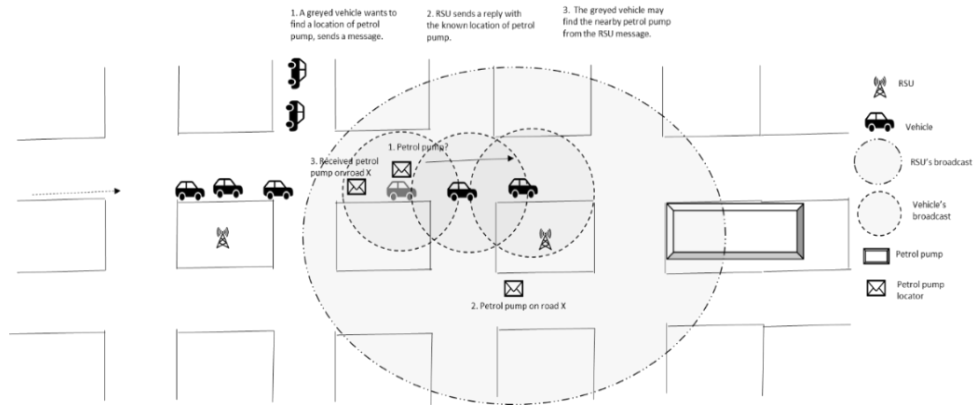

Figure 24. Process of Discovering the Nearest Petrol Pump Station

## 3.8. Announcement of Road Defects

When a vehicle detects a road defect, it broadcasts a message about the issue. The RSU forwards this information directly to the Traffic Authority (TA), which then contacts the RTA to arrange for the appropriate personnel to address the problem. Once the defect is fixed, the TA sends a message to the RSU stating, "resolved: road defect." The RSU then periodically broadcasts this update multiple times within its coverage area. Additionally, the RSU shares this message with nearby RSUs, which also retransmit it within their range. As a result, all vehicles within the coverage area are informed about the resolved road defect issue and can safely use the road again. Figure 25 presents a sequence diagram that illustrates the sequence of operations, while Figure 26 depicts the process of addressing road defects.

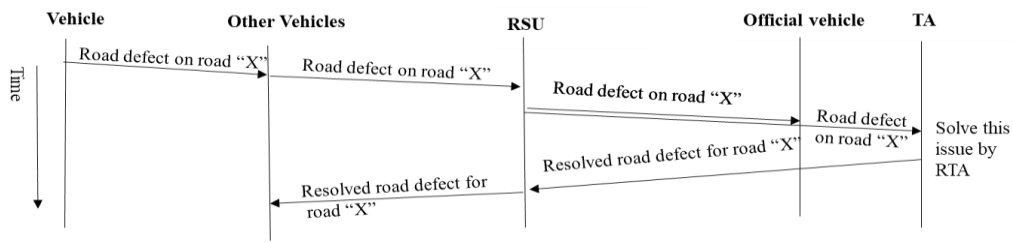

Figure 25. Sequence Diagram for Resolving a Road Defect

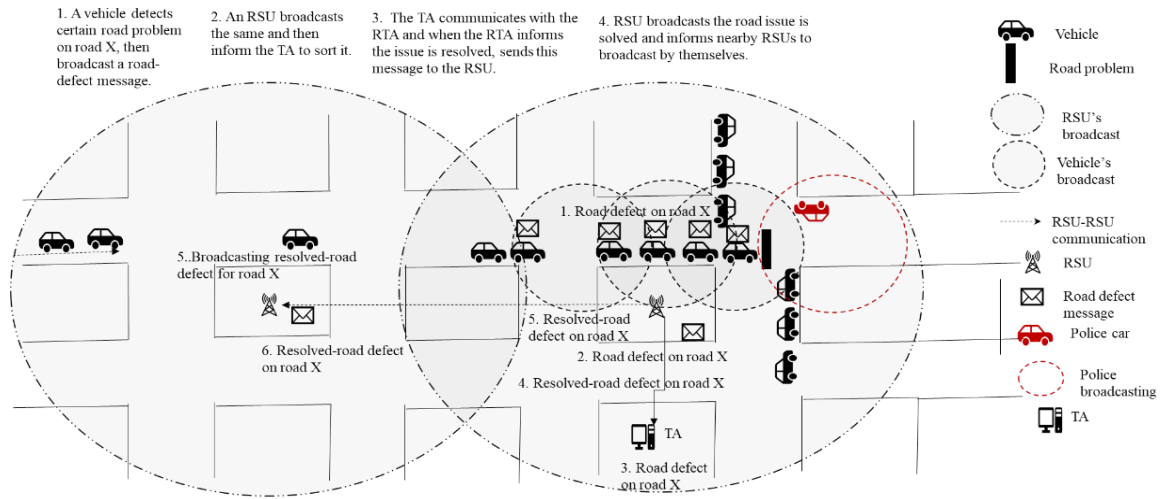

Figure 26. Process of Resolving a Road Defect in VANET

### 3.9. Announcement of Flooding

When a vehicle notices a road is flooded, it sends out a message reporting the issue. The RSU, upon receiving this message, forwards it to the Traffic Authority (TA), which then mobilizes the appropriate personnel to resolve the problem. Once the "flood on road X "is cleared, the TA sends a message to the RSU confirming, "resolved: flood on road X." The RSU subsequently retransmits this update periodically within its coverage area. The RSU also forwards this message to nearby RSUs, which then retransmit the update within their coverage areas. As a result, all vehicles within the range are informed about the cleared flood, allowing them to safely use the road again. Figure 27 illustrates the sequence diagram for this process, while Figure 28 shows the procedure for broadcasting a flood event on the road.

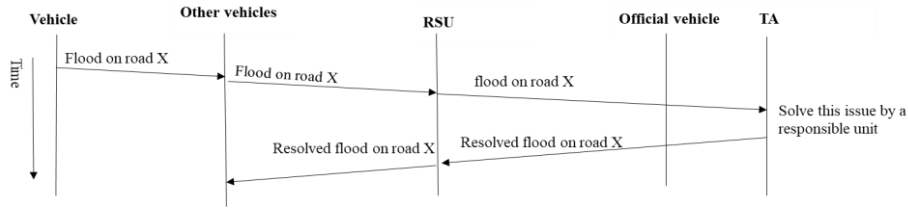

Figure 27. Sequence Diagram for Reporting Flood on a Road Message in VANET

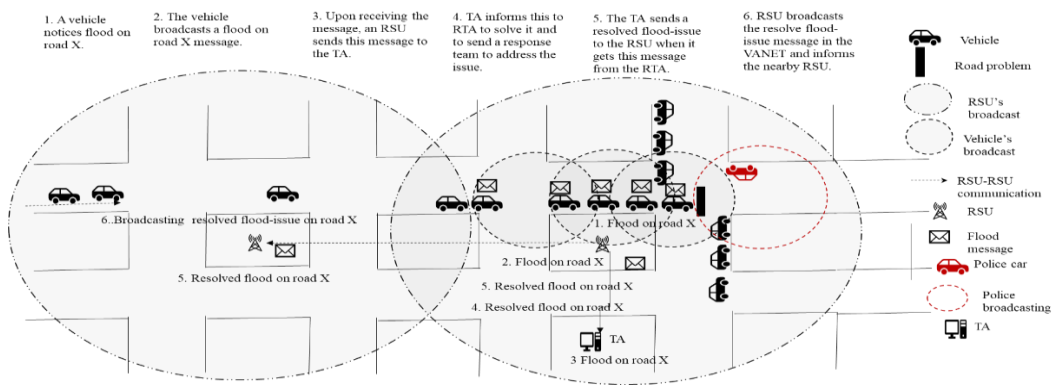

Figure 28. Process of Announcing Flood on a Road Message in VANET

International Journal of Wireless & Mobile Networks (IJWMN), Vol.17, No. 2, April 2025

### 3.10. Announcement of Traffic Signal Malfunction

When a vehicle notices a traffic signal malfunction on road X, it broadcasts a message reporting the issue. The RSU forwards this message to the Traffic Authority (TA), which then arranges for the appropriate personnel to address the problem. Once the traffic signal issue on road X is resolved, the TA sends a message to the RSU saying, "traffic signal problem resolved on road X." The RSU then periodically retransmits this update within its coverage area. Furthermore, the RSU sends the message to nearby RSUs, which also broadcast the resolution within their ranges. As a result, all vehicles within the coverage area are informed about the traffic update and can safely use the road again since the issue has been resolved. Figure 29 illustrates the sequence diagram for broadcasting a traffic signal malfunction event, while Figure 30 shows the dissemination of a traffic signal problem.

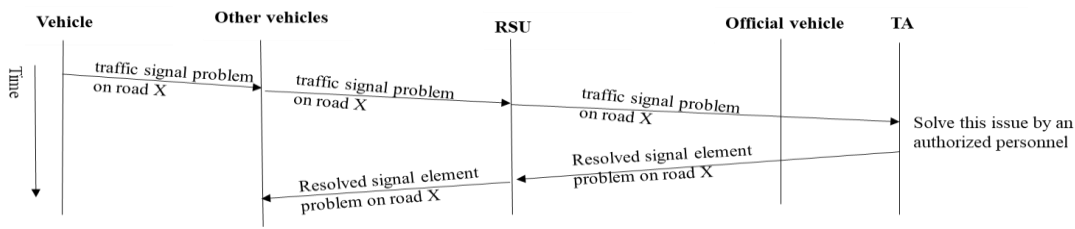

Figure 29. Sequence Diagram for Announcing a Traffic Signal Malfunction

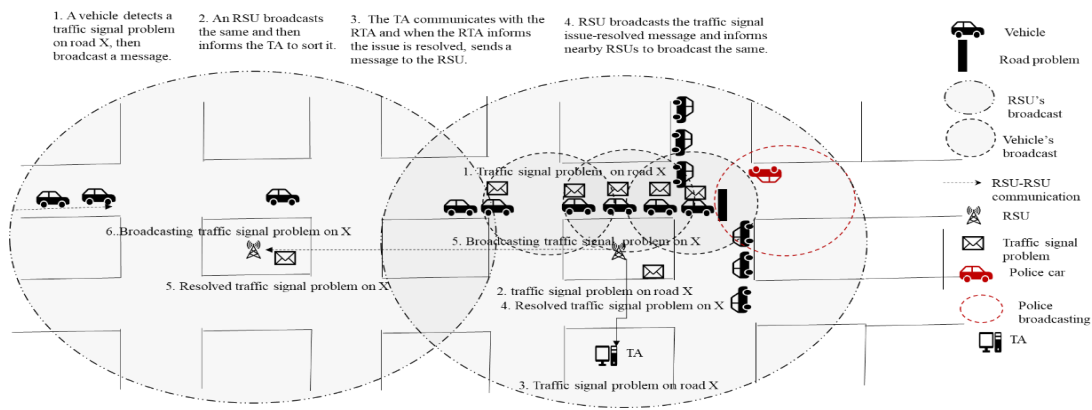

Figure 30. Process for Reporting a Traffic Signal Malfunction

## 4. ANALYSIS OF MESSAGE VOLUME FOR ACCIDENT SCENARIOS

### 4.1. Implementation

We implemented different road traffic scenarios in the VEINS simulator which is composed of OMNet++ and SUMO. This simulator comes with an Erlangen city map for road traffic simulations. In the Erlangen city map, we have placed ten RSUs and one TA to manage events during the experiments. In this map, we have created a circular repeated route where vehicles move during the simulation. The route consists of two lanes and there are some intersections. Also, vehicles follow one after another using the car-following model. The transmission range is set to 300 meters radius as with this range, a smaller number of packet errors occur due to collisions or out-of-range. A vehicle flow is introduced from a fixed location on the map. We consider both sparse and dense traffic conditions using 19 to 139 vehicles. The simulation runs for 1500 simulation seconds. The simulation experiences a warm-up period of 500 simulation seconds, during which no traffic incident is considered, or no vehicle sends any message. This warm-up period is set long enough to allow all vehicles to enter the simulation. The duration is set in such a way so that all the sequences of messages can be exchanged during the period, and we can capture the number of message transmissions required for single-traffic incident management. We repeat each experiment five times to record the average number of messages required for each traffic incident which we define as the average communication overhead for a single traffic incident.

When the simulation begins vehicles enter the simulation area and run with their set speed. A vehicle sends an accident message, which is then relayed by neighbouring vehicles to reach distant vehicles and RSUs. When this message reaches an RSU it coordinates to resolve the situation. It contacts police/ambulance/fire service vehicles as needed. Also, official vehicles relay messages from vehicles and RSUs, as stated in the Section 3. Incident information should go beyond the event area so that vehicles from nearby zones do not enter the event area. Thus, we relay messages up to four hops or when they meet a specific time limit. We consider an accident event on a road in the presence of regular vehicles only and in another set of experiments, both regular and police vehicles participate. Then we measure the average number of messages exchanged from the experiments. This shows on average how many messages are required to manage a single incident considering some scenarios.

### 4.2. Scenario 1

In the first set of experiments, only regular vehicles participate the simulation. We classify the experiments based on how long the relaying will be conducted. We set two criteria to achieve this goal. In the first set of experiments, messages are relayed until the number of hops reaches four and in the second, they are relayed until when their generation time (message freshness) reaches 60 seconds. During the simulation, vehicles stop relaying a message when the condition of relaying becomes false. For both scenarios, vehicle V17 sends an accident message on a road at 550 seconds. This message is then relayed by neighbouring vehicles and RSUs until the relaying limit is met. When an RSU receives an accident message, it sends an "avoid road" message to neighbouring vehicles. The RSU also warns other neighbouring RSUs about the incident to prevent more vehicles from entering the affected area, thereby avoiding further complications. When an RSU receives such a warning from another RSU, it sends the "avoid road" message to its neighbouring vehicles. In these series of simulations, accident, avoid road, and road clear messages are relayed either up to a four-hop distance or when the message freshness becomes 60 seconds. Finally, when the road is cleared, the RSU notifies neighbouring vehicles and RSUs, allowing vehicles from other regions to use the cleared road again.

During the simulation, we measure the number of message transmissions for an accident incident as depicted by the process in Figures 3, 4, 5, 6, and 7. We record the number of messages transmitted in each trial. We run each experiment for five trials to collect the number of messages considering the same number of vehicles running on the road. The number of messages exchanged is determined using the following way: The communication overhead counts the repeated announcements of the accident message from vehicles, and avoid road, road clear messages from RSUs. When an RSU first receives an accident message, it announces the message thrice. If the message is received from another RSU, it retransmits the message twice. An RSU also broadcasts an avoid road message whenever it receives an accident message thrice. Additionally, when an RSU receives an accident message from another RSU, it sends the avoid road message twice. When an RSU obtains an accident message from a vehicle it rebroadcasts the accident notification and additionally, it announces the avoid road message three times. When an RSU receives an avoid road message from another RSU, it repeats the same thrice. When it is

36

received from a vehicle, it relays the same twice. Furthermore, if an old accident message is received again from a vehicle and the incident remains unresolved, an RSU relays the message twice. When the situation is resolved, an RSU which coordinates the situation sends a road clear message so that the vehicles can use the incidental road again. The RSU sends this message to neighbouring RSUs as well so that they can announce the cleared road status among the vehicles in their region. A strict rule is applied in the VANET which states that no vehicle can relay the same message twice as they are already informed about the road condition. This entire process is illustrated in Figures 3, 4, 5, 6, and 7, although the exact number of times each message is broadcasted is not specified in these figures.

### 4.2.1. Discussion

Figure 31 illustrates the average number of messages exchanged among the participant vehicles for both 4-hop and 60-second relaying conditions. Overall, the chart demonstrates an upward trend in communication overhead as the number of vehicles rises. Notably, the 4-hop relaying experiences a higher volume of messages exchanged compared to the 60-second relaying. In the 4-hop relaying scenario, each vehicle that receives a message attempts to relay it to others until the Time to Live (TTL) reaches 4. For each TTL value, all vehicles continue relaying the message, expanding the reach with each successive hop. As the TTL increases, additional vehicles can relay the message, leading to a greater number of messages exchanged compared to the 60-second relaying. Conversely, in the 60-second relaying scenario, vehicles that receive the message will only relay if the message is still within the 60-second time frame. Within 60 seconds, if messages arrive more than once, vehicles do not relay the same message twice, which limits the spread of the message.

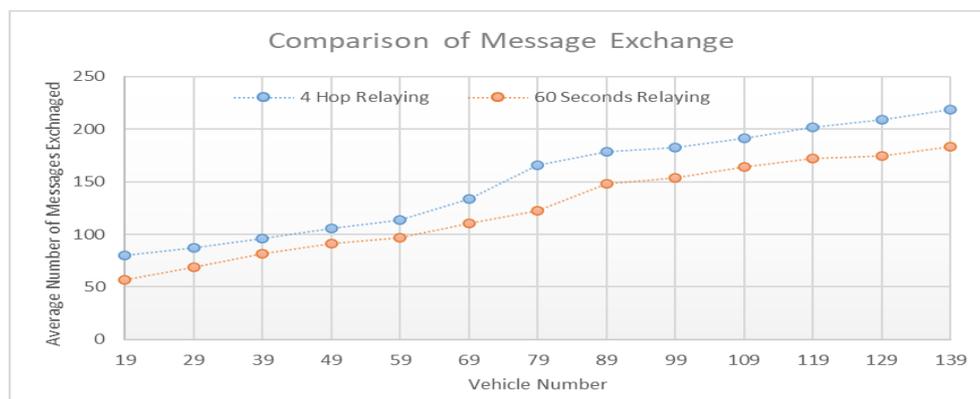

Figure31. Comparison of Communication Overhead of Four-Hop vs. 60-Second Relaying

Although relaying is on for 60 seconds, the 60-second relaying exhibits a smaller number of messages exchanged than the 4-hop relaying in all cases. This means, the 4-hop relaying method enables more vehicles to be informed and leads to a greater number of messages transmitted compared to the 60-second relaying method. In all situations, when a vehicle receives an earlier message, it does not relay the same message again.

### 4.3. Scenario 2

In the second set of experiments, we deployed two police vehicles alongsideregular vehicles. The relaying condition is kept similar during these experiments. We consider messages from the police vehicles as high-priority messages. They are relayed unconditionally as they only cover the simulated area, and this continues until the accident event is sorted. We then measure the number of message transmissions among the vehicles and RSUs in five different trials for 4-hop and 60-second relaying of an incident and event management messages. After this,the average is computed from the number of message transmissionsfrom each trial. Thisaverage communication overhead is used in the chart shown in Figure 32. The series of simulations is conducted using 21 to 131 vehicles including the two police vehicles.

In this series of experiments, the message exchange count comes from the messages generated by the police vehicles besides messages originating from the regular vehicles and RSUs. As the police vehicle sends some messages towards RSU and in front vehicles, they are relayed on the road until the accident is resolved. First, a police vehicle sends an "addressing the incident" message towards a nearby RSU to announce that the police vehicle is approaching to resolve the situation. In response, an RSU sends a confirmation message "OK" to the police vehicle. Furthermore, the police vehicle sends a "free road" message towards the in-front vehicles so that it gets the unobstructed road to reach the incidental area faster. The police vehicle also announces the "cleared road" message when the situation is resolved. Regular vehicles relay the messages as announced by the police vehicle.

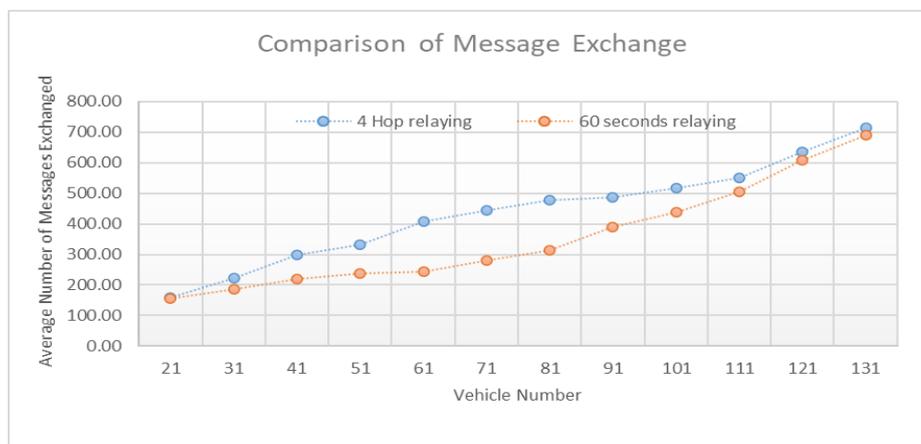

Figure 32. Comparison of Communication Overhead of Four-Hop vs. 60-Second Relaying

### 4.3.1. Discussion

Figure 32 illustrates the average number of message transmissions for both 4-hop and 60-second relaying. It is clear from the chart that more messages are relayed in 4-hop relaying rather than the 60-second relaying experiments. That means many vehicles are informed about the incident with the four-hop relaying case. For low and high numbers of vehicles, the average number of messages exchanged is approximately the same in both conditions. However, with moderate traffic densities, more vehicles are reached with the 4-hop relaying than for 60-second relaying.

## 5. CONCLUSION

This paper presents a traffic incident management model by detailing different sequences and number of messages to manage incidents from the beginning to the clearance of the road. It provides a way to disseminate information about the status of roads, official vehicle involvement, and RSU interaction with users. Having these messages, drivers can avoid or use roads based on their status. We believe this type of framework can be used with a trust/security model or can be used as a separate information dissemination model to instruct or inform drivers about roads and associated conditions. We have created an environment where we configure the number of times a message is broadcasted. We believe for real-world scenarios, the periodicity of the messages and their limits should be set based on the traffic patterns. In the future, we aim to explore the trade-off between the message overhead and the dissemination efficiency within VANETs.

## AUTHORS

**Rezvi Shahariar** received his B.Sc. degree and an M.S. degree in Computer Science from the University of Dhaka, Bangladesh in 2006 and 2007. After some time as a Lecturer and then as an Assistant Professor, he is now an Associate Professor at the Institute of Information Technology, University of Dhaka. He obtained a PhD on trust management for VANETs from Queen Mary, University of London (QMUL). His research interests include wireless network analysis with an emphasis on trust, and security in VANETs.

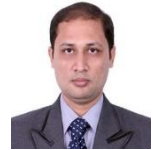

**Chris Phillips** (MIEEE) received a BEng. Degree in Telecoms Engineering from Queen Mary, University of London (QMUL) in 1987 followed by a PhD on concurrent discrete event-driven simulation, also from QMUL. He then worked in industry as a hardware and systems engineer with Bell Northern Research, Siemens Roke Manor Research and Nortel Networks, focusing on broadband network protocols, resource management and resilience. In 2000 he returned to QMUL as a Reader. His research focuses on management mechanisms to enable limited resources to be used effectively in uncertain environments.

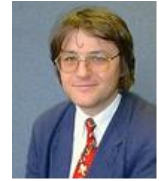